\documentclass[twocolumn]{emulateapj09}

\usepackage{graphicx}
\usepackage{epsfig}

\shorttitle{The Orientation of Sgr~A$^*$}
\shortauthors{Psaltis et al.}

\newcommand{\sgra}{Sgr~A$^*$}

\begin{document}


\title{Event-Horizon-Telescope Evidence for Alignment of the Black
  Hole in the Center of the Milky Way with the Inner Stellar Disk}

\author{Dimitrios Psaltis\altaffilmark{1}, Ramesh
  Narayan\altaffilmark{2}, Vincent L.\ Fish\altaffilmark{3}, Avery E.\
  Broderick\altaffilmark{3,4}, Abraham Loeb\altaffilmark{2}, and
  Sheperd S.\ Doeleman\altaffilmark{2,3}}

\altaffiltext{1}{Astronomy Department,
University of Arizona,
933 N.\ Cherry Ave.,
Tucson, AZ 85721, USA}

\altaffiltext{2}{Harvard-Smithsonian
  CfA, 60 Garden Street, Cambridge, MA 02138, USA}

\altaffiltext{3}{Massachusetts Institute of Technology, Haystack
  Observatory, Route 40, Westford, MA 01886, USA}

\altaffiltext{3}{Perimeter Institute for Theoretical Physics, 31
  Caroline Street North, Waterloo, ON, N2L 2Y5, Canada}

\altaffiltext{4}{Department of Physics and Astronomy, University of
  Waterloo, 200 University Avenue West, Waterloo, ON, N2L 3G1, Canada}

\email{dpsaltis@email.arizona.edu; rnarayan@cfa.harvard.edu;
  abroderick@perimeterinstitute.ca}

\begin{abstract}
Observations of the black hole in the center of the Milky Way with the
Event Horizon Telescope at 1.3~mm have revealed a size of the emitting
region that is smaller than the size of the black-hole shadow. This
can be reconciled with the spectral properties of the source, if the
accretion flow is seen at a relatively high inclination
(50$^\circ$-60$^\circ$). Such an inclination makes the angular
momentum of the flow, and perhaps of the black hole, nearly aligned
with the angular momenta of the orbits of stars that lie within
$\simeq 3$~arcsec from the black hole. We discuss the implications of
such an alignment for the properties of the black hole and of its
accretion flow. We argue that future Event-Horizon-Telescope
observations will not only refine the inclination of \sgra\ but also
measure precisely its orientation on the plane of the sky.
\end{abstract}


\section{INTRODUCTION}

Observations of tight correlations between the properties of galaxies
and of their central, supermassive black holes (e.g., Ferrarese \&
Merritt 2000; Gebhardt et al.\ 2000) provide strong evidence that
their formation and growth histories are tightly coupled (see, e.g.,
Kormendy \& Ho 2013, for a recent review). This connection is further
strengthened by the observed similarity between the histories of star
formation and the evolution of quasars (cf.\ Madau et al.\ 1996 and
Boyle \& Terlevich 1998) as well as by the connection between starbursts
and AGN activity (Kauffmann et al.\ 2003).

An important element in modeling the evolution of supermassive black
holes is the orientation of their spins with respect to the planes of
the galaxies in which they reside and to the angular momenta of the
material that accretes on them (Dotti et al.\ 2013 and references
therein). For example, black holes are believed to acquire most of
their spins via accretion (Berti \& Volonteri 2008; Barausse 2012) and
the magnitude of this effect depends on the relative alignment of the
accretion flow with the black-hole spin. During mergers, black holes
may receive substantial kicks depending on the relative orientation of
the two merging objects (e.g., Baker et al.\ 2008). Finally, the
interaction of the spin of a supermassive black-hole with the
accretion flow (Scheuer \& Feiler 1996; Martin et al.\ 2007; see also
McKinney et al.\ 2013) or with the stars in its vicinity (Merritt \&
Vasiliev 2012) leads to an alignment of the black hole with the inner
regions of the galaxy and further affects the outcome of the above
phenomena.

Observationally, the alignment of a black-hole spin with the axis of a
galactic disk can be inferred primarily by indirect means. Statistical
comparisons of the orientations of host galaxies with the AGN type of
the black-hole accretion flows they harbor find some evidence for
partial alignment (e.g., Lagos et al.\ 2013).  On the other hand,
individual comparisons of the relative orientations between galactic
disks and radio jets (e.g., Schmitt et al.\ 2002) and between galactic
disks and H$_2$O maser sources (Greenhill et al.\ 2009) argue against
such an alignment. The latter studies suggest that the accretion
flows, which are responsible for launching the radio jets and for
generating the masers, are not aligned with the disks of their hosts
galaxies. However, due to obvious resolution limitations, such studies
cannot address whether or not the inner accretion flows are aligned
with the distribution of gas and stars in the vicinity of the black
holes or with the black-hole spins themselves.
  
The Event Horizon Telescope performs sub-mm VLBI observations of the
inner accretion flows around the black holes in the center of the
Milky Way (\sgra; Doeleman et al.\ 2008) and of M87 (Doeleman et
al.\ 2012). Because of its ability to resolve horizon-scale
structures, it provides unique probes of the inner accretion flows
around these black holes, of the jet launching region in M87, and
potentially of the orientations and magnitudes of the black-hole
spins. In the case of \sgra, the orientations of the stellar orbits
within the central 0.1~pc are also known (e.g., Bartko et al.\ 2009)
and can be used in addressing directly the question of alignment
between the inner stellar disk, the accretion flow, and the black-hole
spin.

In this article, we argue that the measurement of the size of the
emitting region of \sgra\ at 1.3~mm can be reconciled with the
spectral properties of the source only if the latter is viewed at a
relatively high inclination. Because of the small inferred size, the
resulting inclination can be inferred robustly even with the current
limited data and depends very weakly on the assumed thermodynamic
properties of the accretion flow.  Moreover, the inferred angular
momentum of the inner accretion flow around \sgra\ is nearly aligned
with that of the inner stellar disk, even though it is not aligned
with the symmetry axis of the Milky Way. We conclude that either the
black hole is not spinning rapidly or that the angular momentum of the
black hole is also aligned with that of the stellar disk.

\section{The Orientation of the Inner Accretion Flow around \sgra}

\begin{figure}[t]
\includegraphics[width=3.5in]{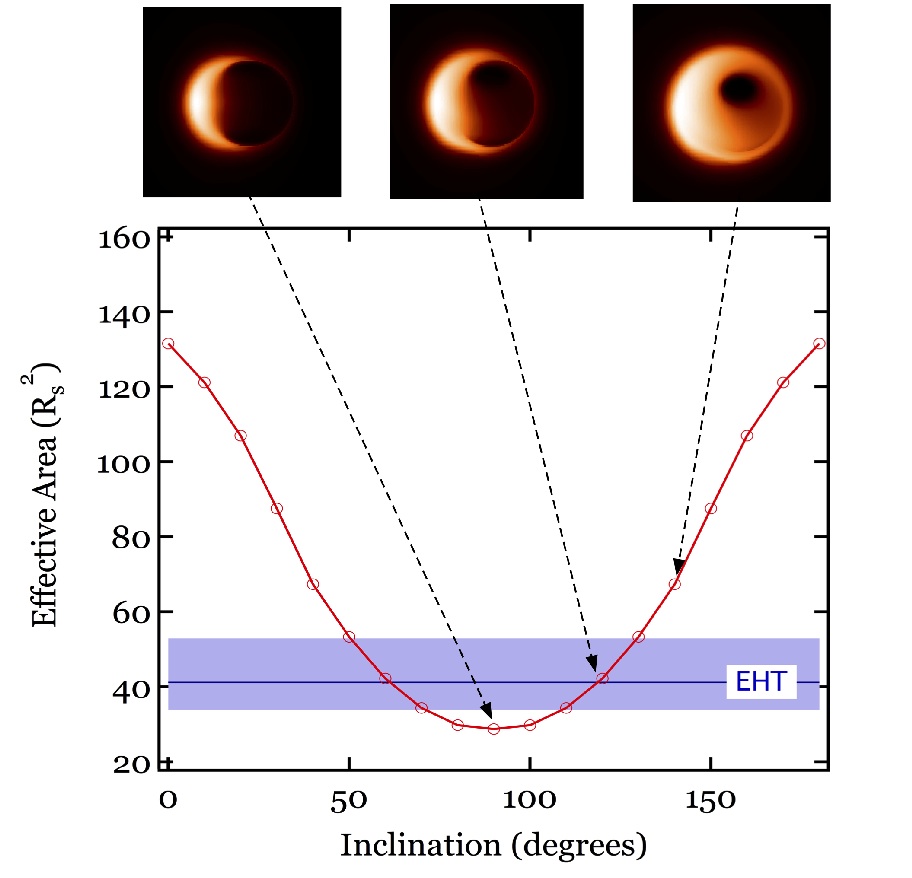}
\caption{The dependence of the effective area of the image of a radiatively
  inefficient flow around \sgra\ on its inclination. All the images
  are calculated with accretion-flow parameters that reproduce, for
  each inclination, the observed sub-mm spectrum of the source. The
  horizontal line shows the measurement and the blue-shaded area the 
  99.7\% uncertainty of effective area reported in 2008 by the
  Event Horizon Telescope. Only for relatively high inclinations, at
  which Doppler effects boost the brightness of the approaching region
  of the flow and cause a large asymmetry in the image, can the
  observed flux from the source be reconciled with its relatively small
  size.}
\label{fig:images}
\end{figure}

Throughout this work, we define the orientation of an angular momentum
vector in terms of two angles: the inclination $\theta$ measured with
respect to the line of sight and the orientation $\phi$ of the
projection of the vector in the sky measured in degrees North of East.
This is the same coordinate system used by Bartko et al.\ (2009) in
defining the orientations of the stellar orbits. The angle $\phi$ is
related to the angle $\xi$ that is measured in degrees East of North
(and that is often used in other articles of the sub-mm image of
\sgra) via $\phi=90^\circ-\xi$.

\subsection{The Orientation Based on EHT Observations}

\begin{figure}[t]
\psfig{file=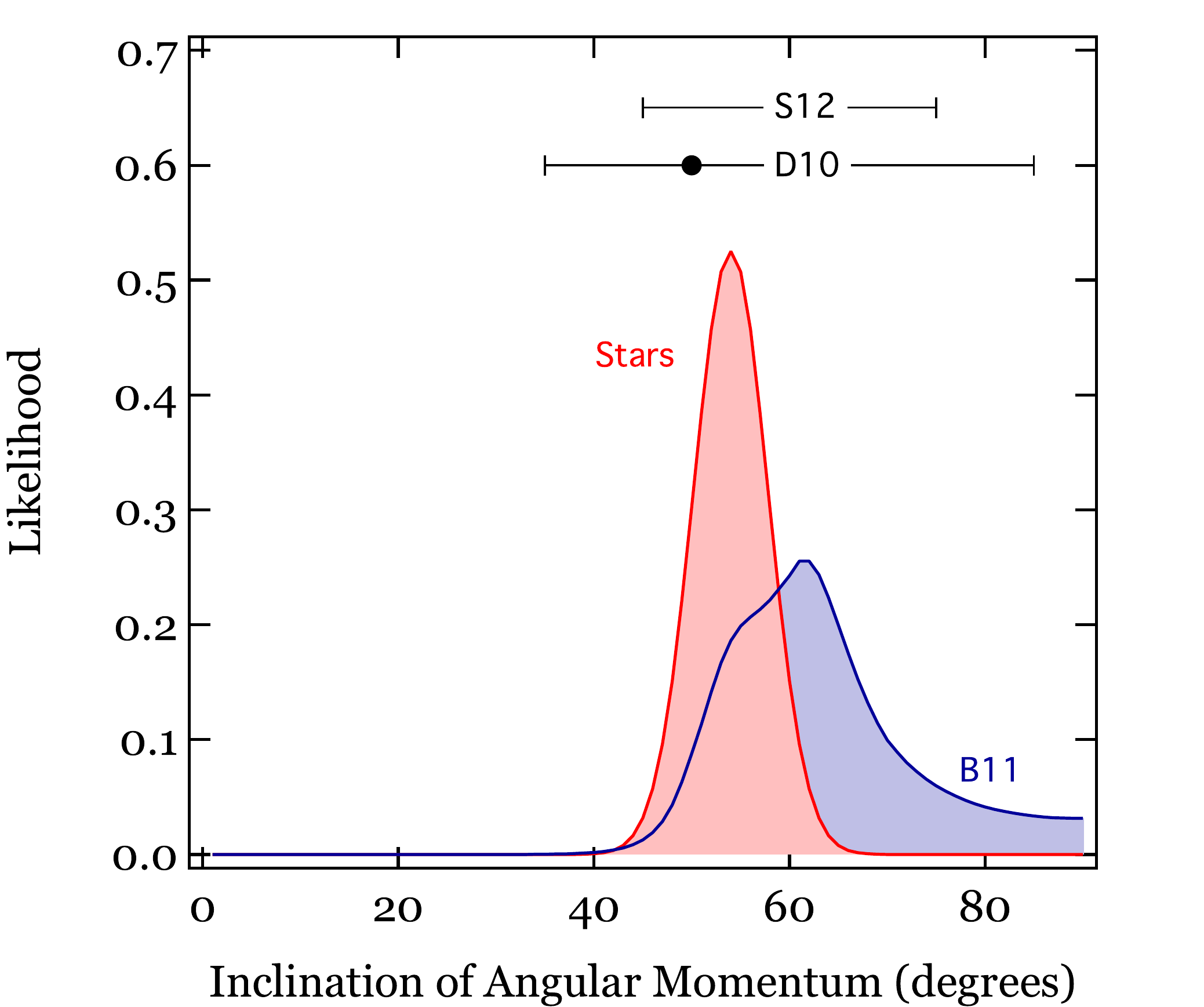,width=3.5in}
\caption{The posterior likelihood of (red) the inclination of the
  angular momentum of the stellar disk within 3.5 arcsec from
  \sgra\ (Bartko et al.\ 2009) and (blue) of the accretion flow, as
  inferred by Broderick et al.\ (2011) when fitting analytical models
  of radiatively inefficient flows to observations made with the Event
  Horizon Telescope. The horizontal error bars above the two curves
  show the most likely value and range for the inclination of the
  accretion flow, as inferred by Dexter et al.\ (2012) and by
  Shcherbakov et al.\ (2012), by fitting numerical GRMHD models to a
  subset of the same Event Horizon Telescope observations.}
\label{fig:inclination}
\end{figure}

Observations of \sgra\ with the Event Horizon Telescope have been
reported for two epochs (2007 and 2009). In each case, the array
resolved a source with a size of $\sim 40~\mu$arcsec, which translates
to about four Schwarzschild radii. This size is consistently smaller
than the expected diameter of the shadow of the black hole (Doeleman
et al.\ 2008; Fish et al.\ 2011), which is about $\sim 5.2$
Schwarzschild radii for zero spin.

The most straightforward way of understanding the fact that the
emitting region in the accretion flow is smaller than the size of the
shadow is to consider the effects of Doppler boosting on the emitted
photons (Doeleman et al.\ 2008). Doppler effects increase the
brightness of the approaching region of the accretion flow and reduce
the brightness of the receding region. This leads to an asymmetric
image with large brightness but very small effective area. In order
for this interpretation to work, however, the accretion flow needs to
be observed at a relatively high inclination.

Figure~\ref{fig:images} shows the dependence on inclination of the
effective area of the accretion flow imaged at 1.3~mm, such that, at
each inclination, the parameters of the accretion model acquire the
value necessary to emit the observed flux. The model images used for
this figure are from Broderick et al.\ (2011a) and the effective area
was calculated, for each inclination, as the intensity weighted angular
area of the image. The blue shaded area shows the area of the image
detected during the original observation with the Event Horizon
Telescope, assuming a circular Gaussian shape (Doeleman et
al.\ 2008). The observed flux and the small size of the emitting
region can be accommodated within this model only if the accretion
flow is observed at a $\sim 60^\circ$ inclination.

Even though the above argument provides a good illustration of the way
in which Doppler effects render the predicted 1.3~mm size of
\sgra\ comparable to the observed one, it is not
conclusive. Figure~\ref{fig:images} compares haphazardly an intensity
weighted area of the crescent-shaped predicted image to the effective
area of a Gaussian that was assumed in fitting the observations. In
fact, a uniform ring of emission, with a size consistent with the
opening angle of the black-hole shadow was also used by Doeleman et
al.\ (2009) to fit the initial EHT data.  Subsequent observations of
\sgra\ with the Event Horizon Telescope allowed for a partial coverage
of the $u-v$ plane, predominantly along E-W baselines. Broderick et
al.\ (2011a) fit an analytical model of the accretion flow to the
observed visibilities and obtained the posterior likelihood for the
inclination of the angular momentum of the inner accretion flow shown
in Figure~\ref{fig:inclination}.  The most likely values for the
inclination were found to be $\theta_{\rm BH}=68^{+5}_{-20}$ degrees,
where the errors correspond to a 68\% confidence level. This is
consistent with the value inferred in Figure~\ref{fig:images} using
the simpler argument based on the image size. (Note here that the
posterior likelihood shown in Figure~\ref{fig:inclination} peaks at a
slightly different angle, because of the marginalization over
$\phi_{\rm BH}$.) Because of the up-down symmetry of the accretion
model, there is a degeneracy with inclinations that are supplementary
to those shown in Figure~\ref{fig:inclination}.

The measurement of the inclination depends rather weakly on the
accretion model, as long as the flow remains nearly equatorial. The
small size of the emitting region makes the result independent of the
details of the accretion flow at larger scales.  What we observe is
primarily emission from a very localized region of the accretion flow
that lies close to the innermost stable circular orbit and is pointing
towards the Earth. Indeed, a similar analysis by Dexter et al.\ (2010)
and Dexter et al.\ (2012), who used 3D GRMHD models of the accretion
flow that are both time-dependent and highly turbulent, resulted in a
very similar inference for the inclination angle: $\theta_{\rm
  BH}=60\pm 15$~degrees. In similar studies, Huang et al.\ (2007,
2009) estimated an inclination of 45$^\circ$.  Finally, Shcherbakov et
al.\ (2012) obtained a good fit of their GRMHD models to the data for
inclinations in the range $45-75$~degrees, with the lowest
inclinations being preferred, if the black hole is spinning slowly.

Note here that the ranges of inclinations inferred in the above
studies have been marginalized over the inferred orientations of the
projection of the angular momentum vector on the sky.  The latter
quantity is not well determined by the current Event Horizon Telescope
observations, primarily because of the lack of substantial N-S
baselines. Broderick et al.\ (2011a) give $\phi_{\rm
  BH}=142^{+15}_{-17}$ degrees and Dexter et al. (2012) give
$\phi_{\rm BH}=160^{+15}_{-86}$. Because only the amplitudes (and not
the phases) of the interferometric visibilities were used in these
studies, there is a degeneracy between these values and values that
are offset by 180~degrees. Taking into account this degeneracy as well
as the large measurement uncertainties leads to a very poor
determination of the orientation of the angular momentum vector on the
sky. In \S4, we identify the optimal baselines at which a
measurement of the correlated flux densities will allow for a precise
inference of this angle. It is worth emphasizing, however, that the most
likely values of the inclination depend rather weakly on the orientation
of the angular momentum vector on the sky because the size of the image
is determined primarily by Doppler effects, as discussed above.

The most significant assumption in the above argument is related to
the alignment of the angular momentum of the accretion flow with the
spin of the black hole.  Dexter \& Fragile (2013) argued that, if the
black hole is spinning moderately to fast, its angular momentum vector
is misaligned from the angular momentum vector of the accretion flow,
and accretion proceeds via a geometrically thick flow, then the size
of the mm image of \sgra\ will not provide a clear measure of its
inclination.  Instead, the shape of the image will be dominated by
emission from the shocks between differentially precessing fluid
elements. This will cause the apparent size of the emitting region to
be small; as such, it will not provide a good measure of the Doppler
boosts between the approaching and receding regions of the flow.  Even
though the current data are insufficient to distinguish between the
two possibilities, the similarity in the visibility amplitudes between
observations separated by several years (Broderick et al.\ 2011a)
suggests that the mm image of \sgra\ is not dominated by transient
emission, as we would expect from shocked material. Future EHT
observations will have the sensitivity to clearly disentangle such
time-variable structures, if they exist (Doeleman et al.\ 2009; Fish
et al.\ 2009b).

\begin{figure}[t]
\psfig{file=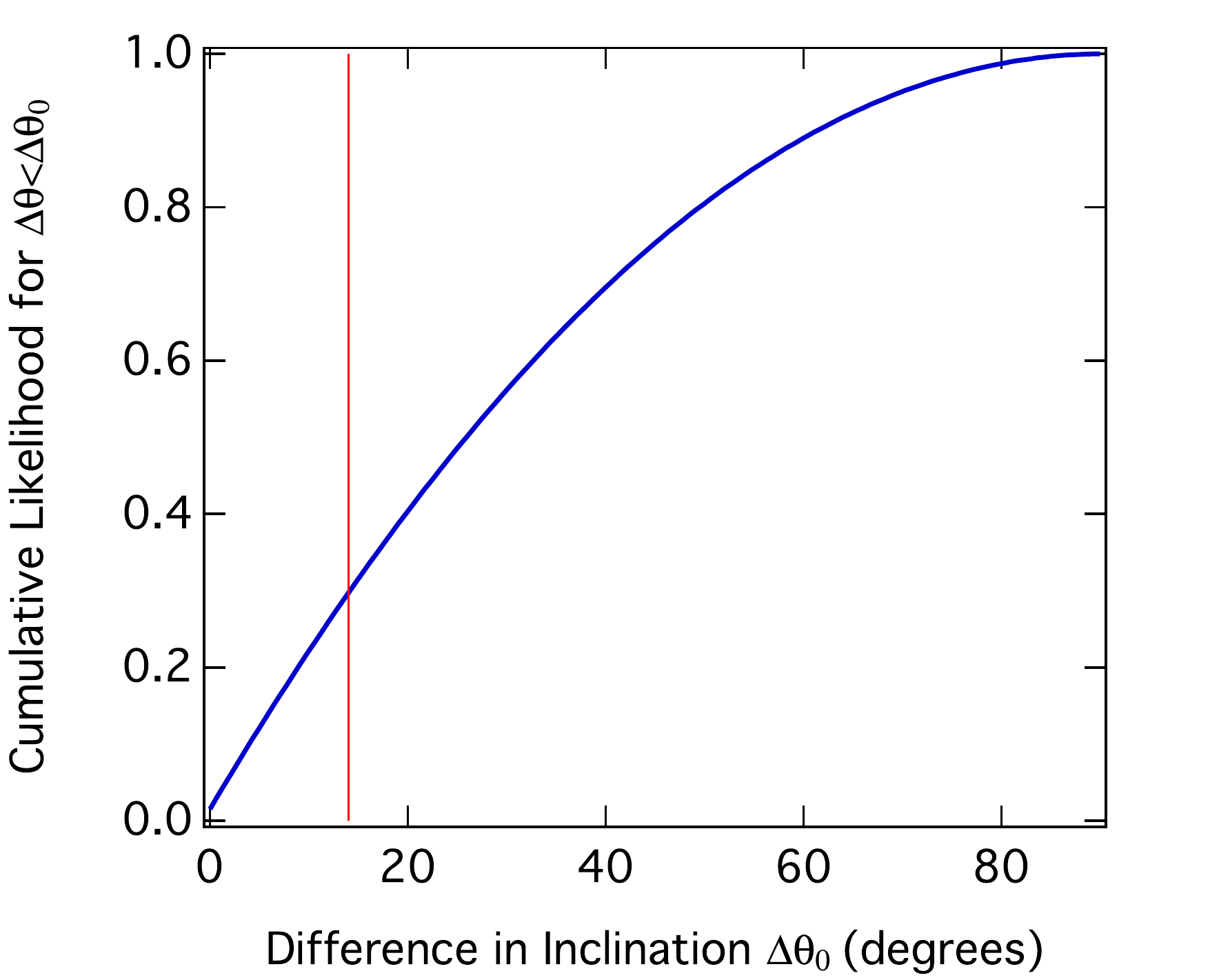,width=3.5in}
\caption{The posterior cumulative likelihood that the orientation of
  the inner accretion flow around \sgra\ (as measured with the EHT)
  and of the inner stellar disk (as inferred from the stellar orbits)
  is within an angle $\Delta\theta_0$, if they are both randomly
  oriented in the sky. The two orientations align to within
  $14^\circ$ of each other by coincidence only $\lesssim 25$\% of
  the time.}
\label{fig:probability}
\end{figure}

\subsection{Alignment of the Accretion Flow with the Stellar Disk}

The orientation and inclination of the inner accretion flow around
\sgra, as inferred in the previous section, is misaligned with that of
the galactic plane, for which $\theta_{\rm MW}\simeq 90^\circ$ and
$\phi_{\rm MW}\simeq 330^\circ$ (Reid \& Brunnthaler 2004). This is
not surprising, however, because the stars within a few arcsec from
\sgra, for which monitoring of their orbits has been possible, appear
to lie predominantly in a series of concentric disks that are inclined
with respect to the galactic plane.

It is widely believed that the majority of stars orbiting the black
hole at projected distances in the range $\simeq 1-10$~arcsec lie on a
clockwise disk (Levin \& Beloborodov 2003; Genzel et al.\ 2003;
Paumard et al.\ 2006; Lu et al.\ 2009; Bartko et al.\ 2009).  There is
evidence that the disk is warped (Bartko et al.\ 2009), with an
orientation that evolves with distance. However, the orbital angular
momentum vectors of the stars with projected distances that are
closest to \sgra\ point, on average, towards a direction characterized
by $\theta_{\rm CW}=54\pm 3.2$~degrees and $\phi_{\rm CW}=256\pm
3.2$~degrees (Bartko et al.\ 2009), with a distribution that has a
HWHM of $\simeq$16~degrees.  A different, counterclockwise stellar
disk at similar projected distances has also been suggested with an
inclination of $\theta_{\rm CCW}\simeq 142^\circ$ and an orientation
of $\phi_{\rm CCW}\simeq 200^\circ$ (Barko et al.\ 2009; see also
Genzel et al.\ 2003; Paumard et al.\ 2006; see, however, Lu et
al.\ 2009; Yelda et al.\ 2014).

Recently, a cloud (G2) was discovered in the vicinity of \sgra
(Gillessen et al.\ 2012), with a trajectory that eventually intercepted
\sgra\ in mid-late 2013 (Gillessen et al.\ 2013). The orbital plane of
the cloud was found to be within the HWHM of the orientations of the
stellar orbits; Gillessen et al.\ (2012) give $\theta_{\rm
  G2}=70.52$~degrees and $\phi_{\rm G2}=264.2$~degrees, whereas Phifer
et al.\ (2013) infer $\theta_{\rm G2}=59\pm 3$~degrees and $\phi_{\rm
  G2}=304\pm 11$~degrees.

\begin{figure}[t]
\psfig{file=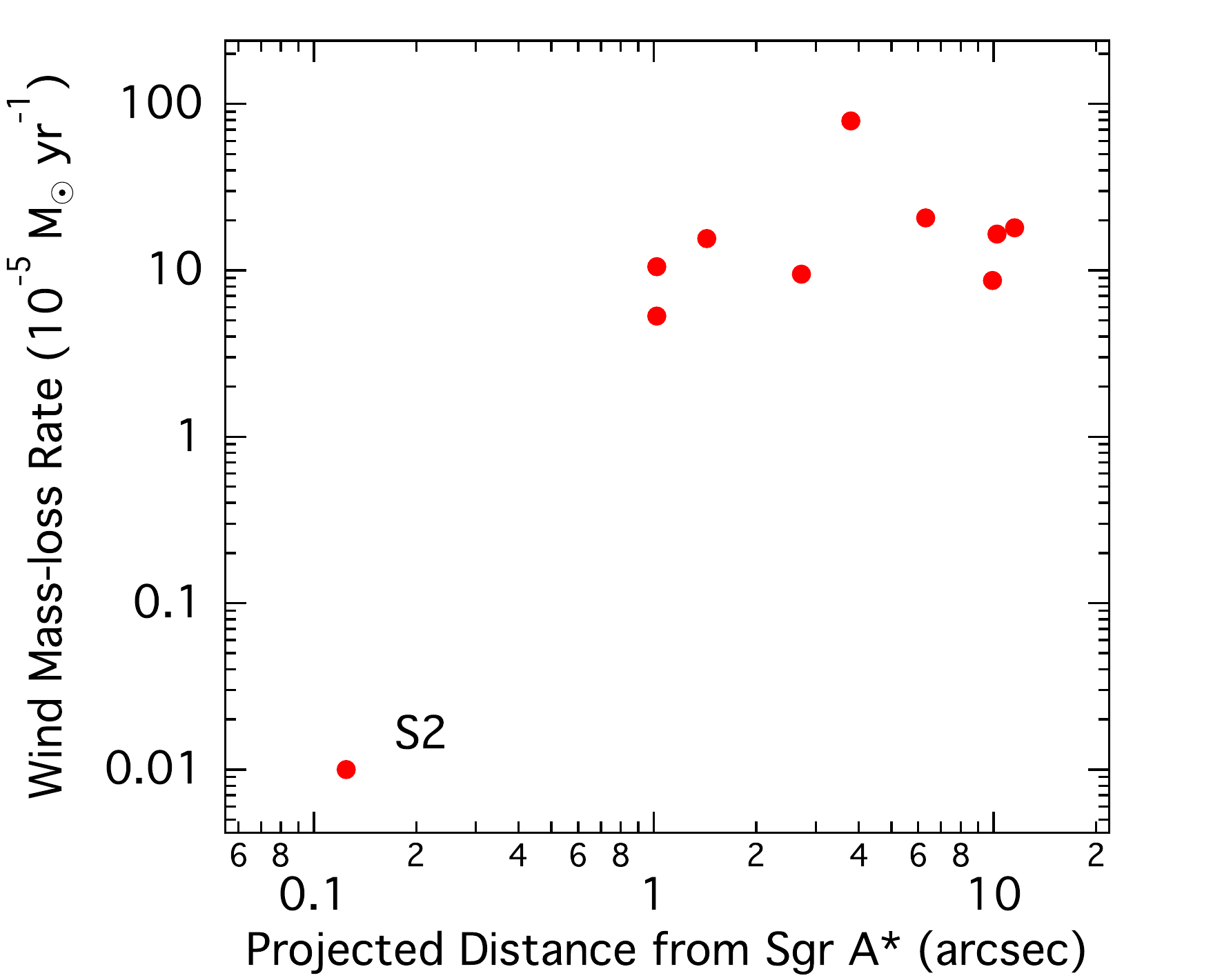,width=3.5in}
\caption{The inferred wind mass-loss rates for different stars in
  orbit around \sgra (from Najarro et al.\ 1997; Martins et al.\ 2007,
  2008). The stars at angular distances $\simeq 1-3$~arcsec, the
  orbits of which appear to be aligned with the inner accretion flow
  around the black hole, have the high mass-loss rates that are
  believed to be supplying the gas to the black hole.}
\label{fig:winds}
\end{figure}

\begin{figure*}[t]
\centerline{
\includegraphics[width=3.2in]{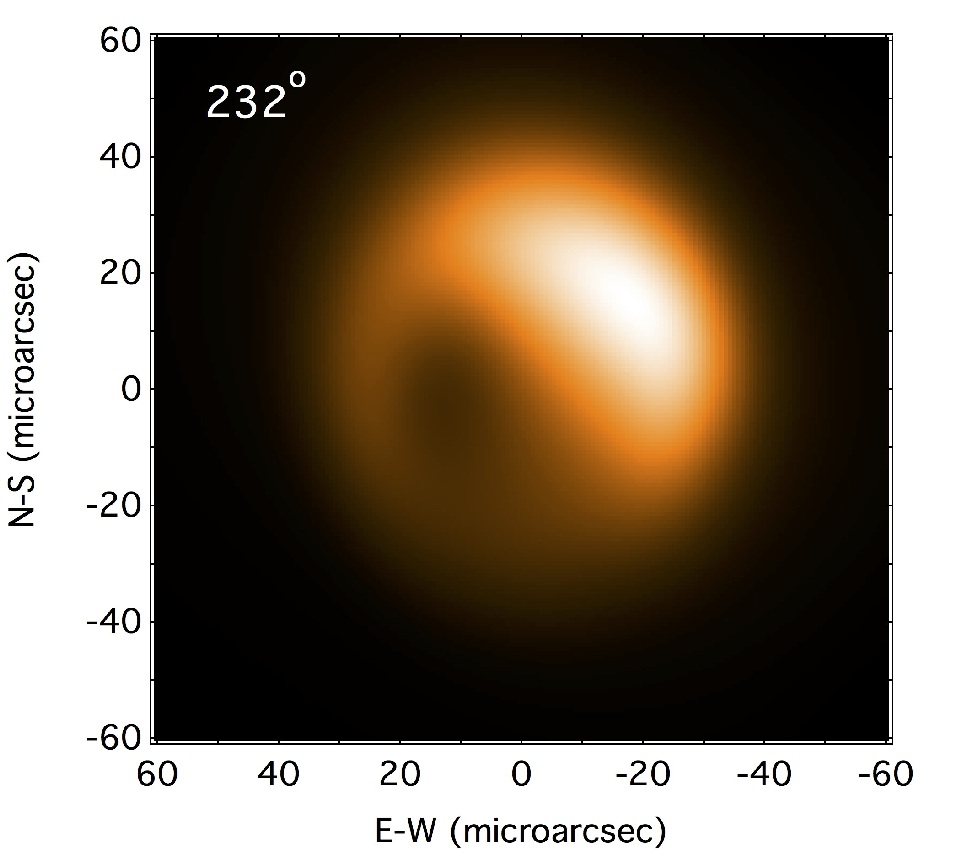}
\includegraphics[width=3.2in]{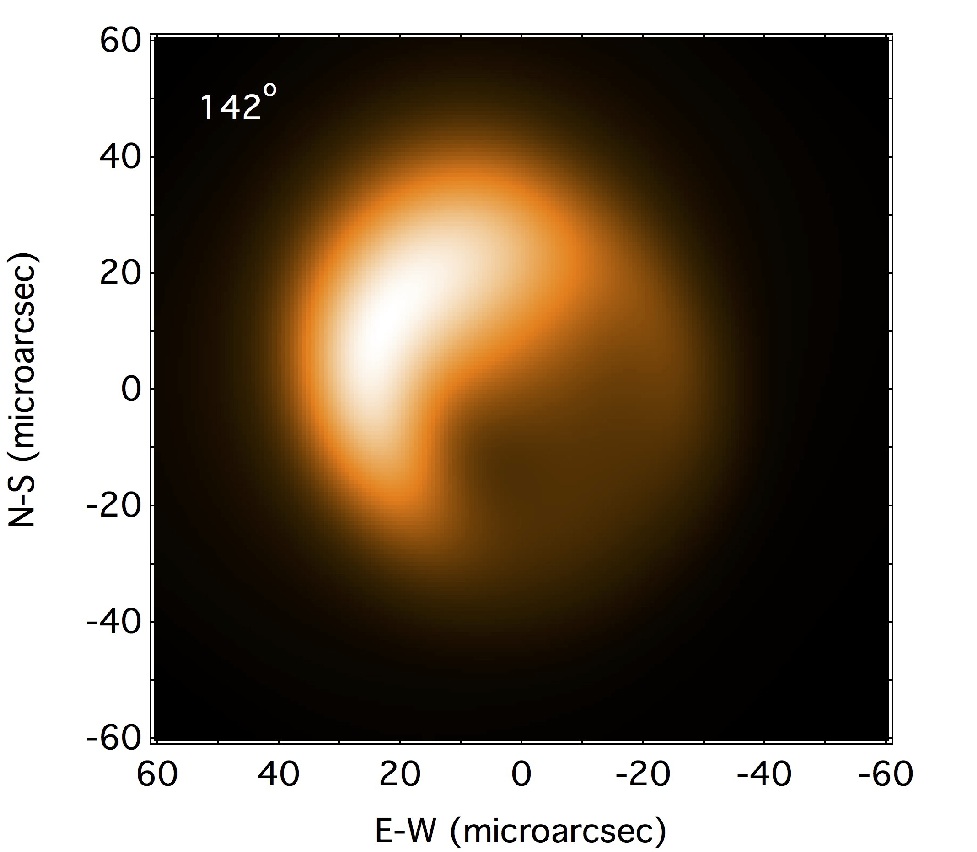}}
\centerline{
\includegraphics[width=3.2in]{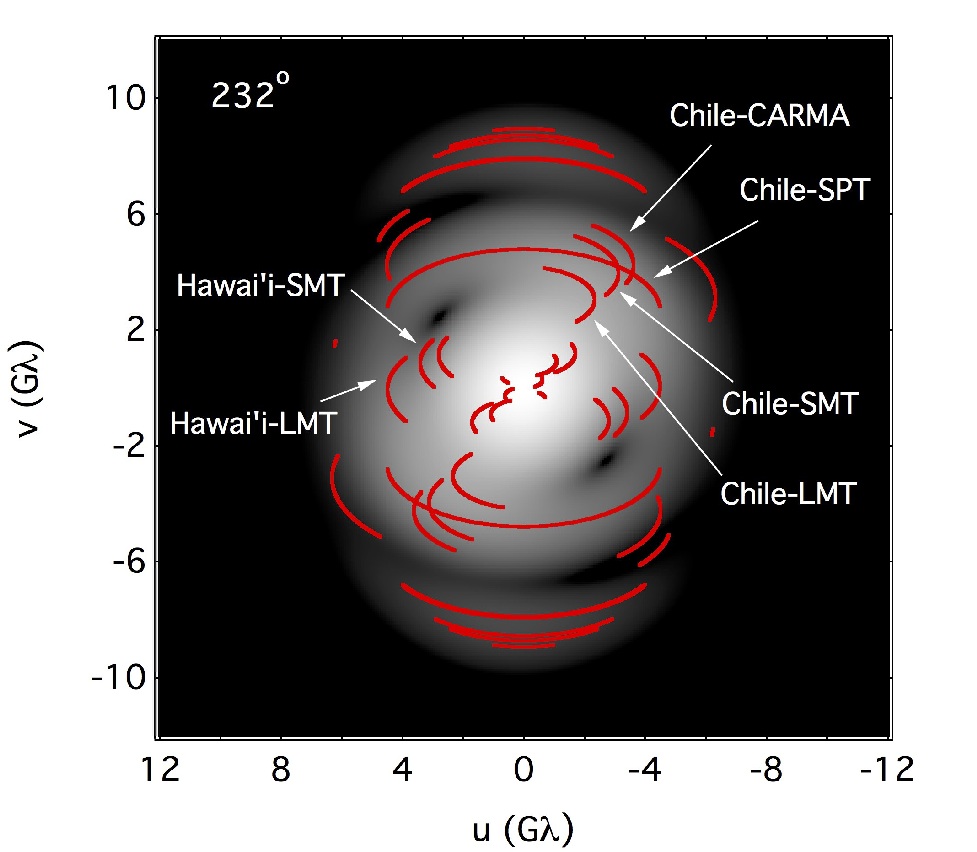}
\includegraphics[width=3.2in]{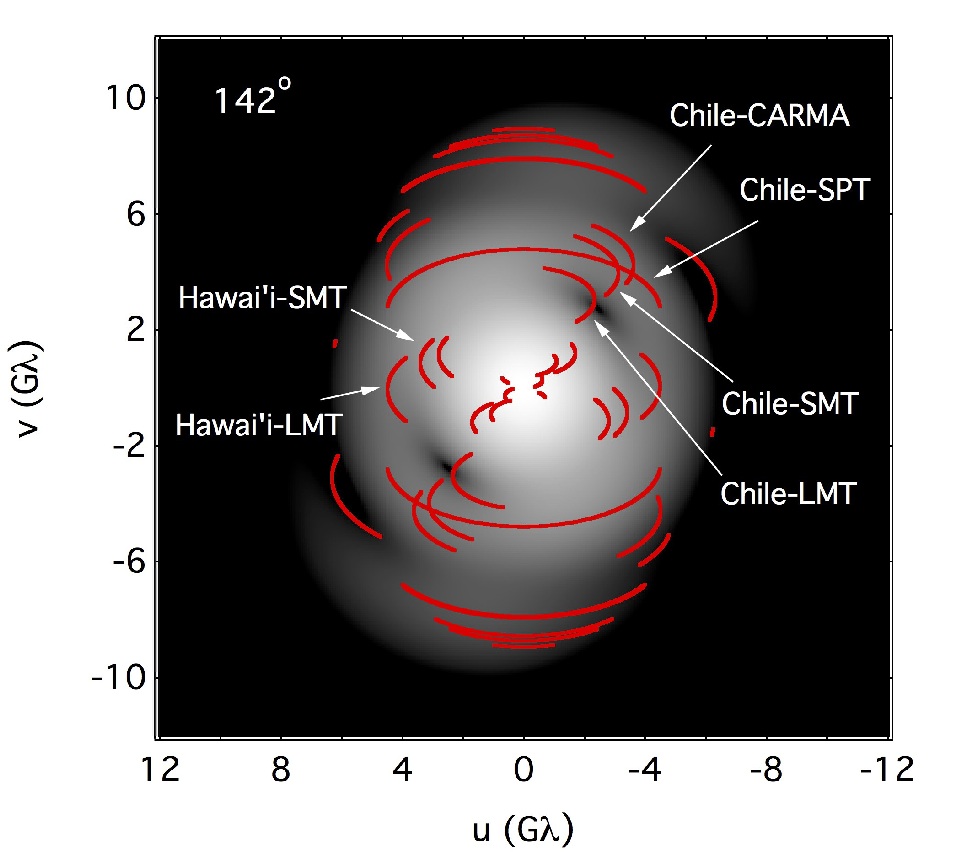}}
\caption{Simulated scattering-broadened images (top panels) and
  $u-v$ maps (bottom panels) of the accretion flow around \sgra\ at
  1.3~mm, for an inclination of $60^\circ$ and for two different
  orientations in the sky.  In the $u-v$ maps, the tracks of various EHT
  baselines are overplotted.  Distinguishing between different
  orientations can be most easily achieved by combining data on the
  Chile-LMT and Hawai'i-SMT baselines, which probe the nulls in the
  visibilities associated with the black-hole shadow.}
\label{fig:uv}
\end{figure*}

\begin{figure}[t]
\psfig{file=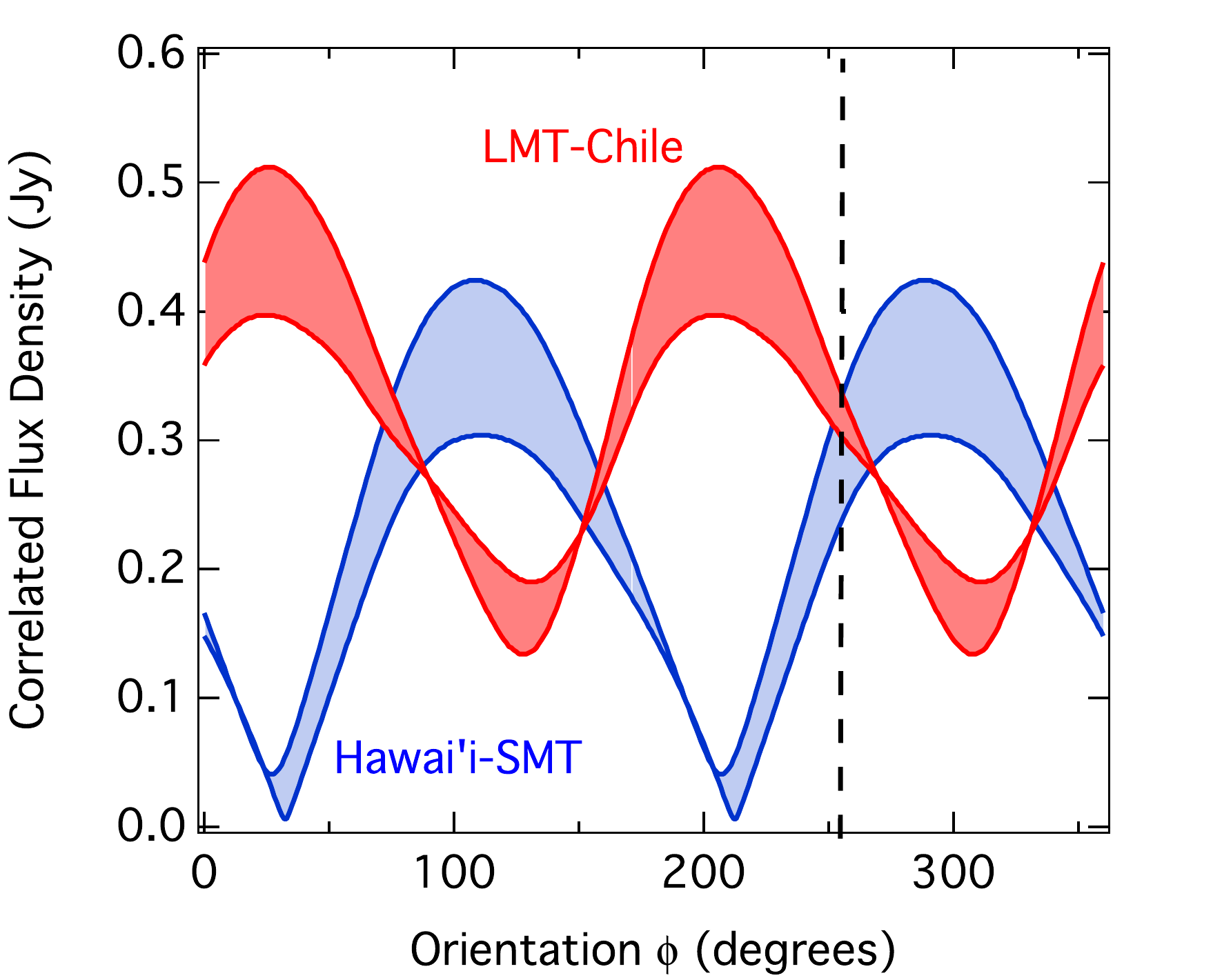,width=3.5in}
\psfig{file=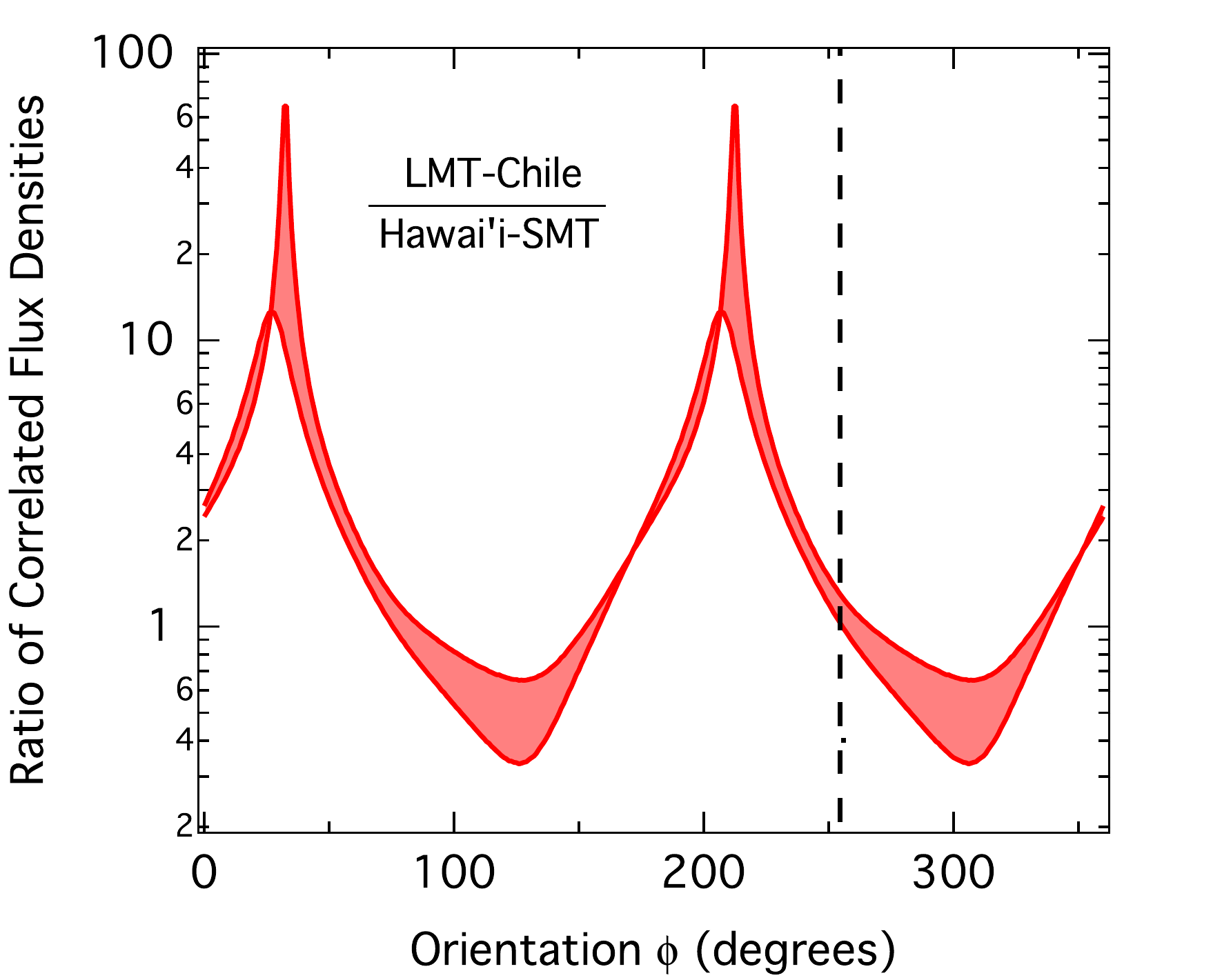,width=3.5in}
\caption{Predicted correlated flux densities for the LMT-Chile and
  Hawai'i-SMT baselines {\em (top)\/} and their ratios {\em
    (bottom)\/} as a function of the orientation of the inner
  accretion flow. The width of each curve shows the variation of the
  predicted flux densities with inclination in the $50-60^\circ$
  degree range. The vertical dashed line shows the orientation of the
  inner stellar disk.}
\label{fig:visibility}
\end{figure}

Figure~\ref{fig:inclination} compares the posterior likelihoods for the
inclinations of the stellar disk to that of the inner accretion flow
around \sgra, as inferred by EHT observations. This figure suggests
that the angular momentum of the inner accretion flow may be
aligned with the angular momenta of the stars in the inner disk. 

Figure~\ref{fig:probability} shows the cumulative posterior likelihood
that the inclinations of the two angular momentum vectors are within
an angle $\Delta\theta_0$ by coincidence. In order to take into
account the Lutz-Kelker bias (since the inclination angle can only be
positive by definition), we generated a large number of Monte Carlo
realizations of the two vectors with random directions in the sky. We
then assigned Gaussian errors in the measurements of the inclinations
of the two vectors, with a dispersion of $15^\circ$ degrees for the
black hole and of $3^\circ$ for the stellar disk, to match the
observational errors. We finally calculated the fraction of the
resulting measurements that lie within an angle $\Delta\theta_0$. The
likelihood that the two inclinations are within $14^\circ$ by
coincidence is $\lesssim 25$\%. If we adopt the inclination inferred
by Dexter et al.\ (2012) so that the two inclinations are within
$6^\circ$ of each other, we find this likelihood to be $\lesssim
13$\%.

\section{Discussion}

In the previous section, we compared the inclination of the inner
accretion flow around \sgra, as inferred from observations with the
Event Horizon Telescope, to that of the stellar disk that
lies within the central 3.5~arcsec and found them to be within
$6-14^\circ$ of each other. We then used this evidence to argue that
the angular momentum of the inner accretion flow around the black hole
in the center of the Milky Way is nearly aligned with that of the
stars close to it.

It is worth pointing out that the inner accretion flow appears to be
aligned with the stellar disk 3.5~arcsec away, even though a large
number of stars (the S-star cluster) are closer to the black hole and
have more isotropic orbits (see, e.g., Gillessen et al.\ 2009). This
should not be very surprising, however, given that the stars at
$1-10$~arcsecs have much higher inferred wind mass-loss rates and are
the ones that are probably supplying the accreting material to the
black hole (see Fig.~\ref{fig:winds} and Rockefeller et al.\ 2004;
Cuadra et al.\ 2006).

The lack of variability in the size of the image of \sgra\ (see Fish
et al.\ 2011; Broderick et al.\ 2011a) and the evidence shown in
Figure~\ref{fig:inclination} suggest that the emission from shocks due
to the differential precession of nearby fluid elements is absent from
the image of \sgra. This can be achieved either if the black-hole
angular momentum is nearly aligned with that of the flow or if the
black hole is slowly spinning (Dexter et al.\ 2012 suggest within
$\simeq 15^\circ$ or for $a<0.3$)

The black hole may be aligned with the stellar cluster and the inner
accretion flow either because of spin-orbit exchanges with the stellar
cluster (Merritt \& Vasiliev 2012) or due to dissipation of its
angular momentum in a geometrically thin disk (Scheuer \& Feiler
1996). Merritt (2010) inferred that the inner stellar cluster has a
mass of $\sim 10^4 M_\odot$ and a core radius of $\sim 0.1$~pc. Even
though the orbital angular momentum in this cluster is comparable to
that of the black hole, the mutual precession timescale is
$>10^{10}$~yr (Merritt \& Vasiliev 2012). Such a long timescale makes
it highly unlikely that \sgra\ could have been aligned because of
spin-orbit exchanges with the stellar cluster.

If \sgra\ was accreting at some point at a relatively high rate via a
geometrically thin accretion disk, it would have aligned after a time
of (Scheuer \& Feiler 1996; Volonteri et al.\ 2005)
\begin{eqnarray}
t_{\rm align}&\simeq& 10^7\chi^{2/3}
\left(\frac{\alpha}{0.1}\right)^{5/3}
\left(\frac{\eta}{0.1}\right)\nonumber\\
&&\qquad\qquad
\left(\frac{H}{0.1R}\right)^{2/3}
\left(\frac{\dot{M}}{0.1\dot{M}_{\rm Edd}}\right)^{-1}~\mbox{yr}\;,
\label{eq:al_time}
\end{eqnarray}
where $\alpha$ is the viscosity parameter, $\chi$ is the spin of the
black hole, $H$ is the scale height of the disk at distance $R$, and
$\dot{M}/\dot{M}_{\rm E}$ is the mass accretion rate in units of the
Eddington critical rate. In that amount of time, \sgra\ would have
accreted a large fraction of its own mass and significantly more mass
than is currently available within 0.1~pc, in the stellar cluster.
This is, therefore, a highly unlikely possibility, as well. 

It is worth emphasizing here that relation~(\ref{eq:al_time}) assumes
that the only mechanism that causes the disk angular momentum to align
with the black-hole spin is related to the Bardeen-Petterson effect
(Bardeen \& Petterson 1975). If additional aligning mechanisms are
effective, such as the one discussed in McKinney et al.\ (2013) that
invokes large magnetic torques near spinning black holes with
extensive jets, then the above timescale will be altered. However,
such additional effects, which have comparable strength as the
Bardeen-Petterson torques, will not significantly change the very long
timescale calculated above. The most likely conclusion is that
\sgra\ is slowly spinning, which is also indicated by the analysis
of brightness distribution in the 1.3~mm images of the inner accretion
flow (e.g., Broderick et al.\ 2011a).

Observations of the 1.3~mm image of \sgra\ with the Event Horizon
Telescope in the near future, when more stations are added to the
interferometer, will break the degeneracies in the measurement of the
two orientation angles and provide additional evidence to support or
refute the result presented here.

First, in order for the array to maximize its ability to measure the
orientation of the angular momentum of the accretion flow on the plane
of the sky, a set of orthogonal baselines with a separation comparable
to the projected size of the black-hole shadow must be used. This is
demonstrated in Figure~\ref{fig:uv}, which shows the
scattering-broadened images and the corresponding $u-v$ maps for two
different orientations of the accretion flow that are separated by
$90^\circ$. For this figure, we have used the images of Broderick et
al.\ (2011a) that best reproduce the currently observed spectra,
polarization limits, and size of \sgra.

The black-hole shadow generates two null regions in the $u-v$ plane
that are probed by a number of nearly orthogonal baselines, e.g., the
Chile-LMT baseline in the N-S direction and the Hawai'i-SMT baseline
in the E-W direction. If the black-hole shadow is indeed imprinted on
the image, then visibility amplitudes of these two baselines will be
anticorrelated. Moreover, the ratio of the visibility amplitudes
measured along these two baselines will provide a direct measure of
the orientation of the angular momentum of the accretion flow in the
sky (see~Fig.~\ref{fig:visibility}). In particular, if the inner
accretion flow is aligned with the stellar disk (i.e., if
$\phi=256^\circ$), then the two baselines should have comparable
correlated flux densities that are of order $\sim 0.3$~Jy.

When only the visibility amplitudes of the $u-v$ maps are used, then
there is an additional degeneracy between orientations that differ by
180$^\circ$. This degeneracy can be lifted with the use of closure
phases that have already become available to 1.3~mm observations of
\sgra (see Broderick et al.\ 2011b). Finally, if quasi-coherent
regions of enhanced emission (i.e., blobs) are seen orbiting the black
hole (see, e.g., Broderick \& Loeb 2006; Doeleman et al.\ 2009), then
whether these bright regions appear to move clockwise or
counterclockwise on the image plane will break the degeneracy between
supplementary inclinations.

The orientation of the spin of \sgra\ can be also be inferred if a jet
is discovered that originates from the central black hole. Several
candidate jet structures have been identified during the last decade,
with orientations that practically cover the entire range of
possibilities (see the discussion in Li et al.\ 2013). If the
black-hole spin is aligned with the orbital angular momenta of the
stars in the inner clockwise disk, which themselves are almost
orthogonal to the orientation of the galactic pole, then the jet
structures most recently identified by Li et al.\ (2013) and Su \&
Finkbeiner (2012) cannot be related to \sgra. On the other hand, the
structures identified by Muzi\'c et al. (2007) and Yusef-Zadeh et
al.\ (2012) will be roughly aligned with the inferred spin axis of the
central black hole. The situation is, of course, reversed, if the
orientation of the black-hole spin is at $\sim 142^\circ$, as inferred
by Broderick et al.\ (2011a), using the limited EHT observations that
are currently available. These considerations, albeit currently
inconclusive, demonstrate how combining EHT observations with those of
stellar orbits and jet-like structures in the galactic center can shed
light to the complex interaction between the inner galaxy and its
central black hole.

\acknowledgements

D.P. was supported by NSF CAREER award AST~0746549, NASA/NSF TCAN
award NNX14AB48G, and by NSF award AST~1312034. R.N. was supported in
part by NSF grant AST~1312651. AL was supported in part by NSF grants
AST~0907890 and AST~1312034 and NASA grants NNX08AL43G and NNA09DB30A.
Work at MIT Haystack Observatory was made possible by grants from the
National Science Foundation (NSF) and the Gordon and Betty Moore
Foundation (GBMF-3561).  A.E.B. receives financial support from the
Perimeter Institute for Theoretical Physics and the Natural Sciences
and Engineering Research Council of Canada through a Discovery
Grant. Research at Perimeter Institute is supported by the Government
of Canada through Industry Canada and by the Province of Ontario
through the Ministry of Research and Innovation.

\end{document}